\documentclass[aps,prl,showpacs,twocolumn,groupedaddress]{revtex4}

\usepackage{graphics}

\begin{document}


\title{Microscopic Study of Slablike and Rodlike Nuclei:\\
  Quantum Molecular Dynamics Approach}


\author{Gentaro Watanabe$^{a,b}$, Katsuhiko Sato$^{a,c}$,
  Kenji Yasuoka$^{d}$ and Toshikazu Ebisuzaki$^{b}$}
\affiliation{
$^{a}$Department of Physics, University of Tokyo,
Tokyo 113-0033, Japan
\\
$^{b}$Division of Computational Science, RIKEN,
Saitama 351-0198, Japan
\\
$^{c}$Research Center for the Early Universe, 
University of Tokyo,
Tokyo 113-0033, Japan
\\
$^{d}$Department of Mechanical Engineering, Keio University,
Yokohama 223-8522, Japan}


\date{\today}

\begin{abstract}
Structure of cold dense matter
at subnuclear densities is investigated
by quantum molecular dynamics (QMD) simulations.
We succeeded in showing that the phases with slab-like and
rod-like nuclei etc. can be formed dynamically
from hot uniform nuclear matter
without any assumptions on nuclear shape.
We also observe intermediate phases, which has complicated nuclear shapes.
Geometrical structures of matter are analyzed
with Minkowski functionals, and it is found out that intermediate phases
can be characterized
as ones with negative Euler characteristic.
Our result suggests the existence of these kinds of phases
in addition to the simple ``pasta'' phases
in neutron star crusts.
\end{abstract}

\pacs{21.65.+f, 26.50.+x, 26.60.+c, 02.70.Ns}

\maketitle


Collapse-driven supernovae (SNe)
and the following formation of neutron stars (NSs)
are the most dramatic processes during the stellar evolution.
These objects provide not only astrophysically significant phenomena
but also interesting material phases inside them;
both are strongly connected with each other.
Investigating the properties of nuclear matter
under extreme condition is one of the essential subjects
to clarify the mechanism of collapse-driven SNe \cite{bethe}
and the structure of NS crusts \cite{review}.
This subject is also interesting
in terms of the fundamental problem of the complex fluids of nucleons.

At subnuclear densities,
where nuclei are about to melt into uniform matter,
it is expected that the energetically favorable configuration
possesses interesting spatial structures
such as rod-like and slab-like nuclei and rod-like and spherical bubbles etc.,
which are called nuclear ``pasta''.
This picture was first proposed
by Ravenhall et al. \cite{rpw}
and Hashimoto et al. \cite{hashimoto}.
These works,
which are based on free energy calculations
with liquid drop models assuming some specific nuclear shapes,
clarify that
the most energetically stable nuclear shape is
determined by a subtle balance between nuclear surface and Coulomb energies.
Although detailed aspects of phase diagrams
vary with nuclear models \cite{lorenz},
the realization of the ``pasta'' phases as energy minimum states
can be seen in a wide range of nuclear models and
basic feature of phase diagrams is universal \cite{gentaro1,gentaro2},
i.e. with increasing density,
the shape of the nuclear matter region changes like
\ sphere $\rightarrow$ cylinder $\rightarrow$ slab $\rightarrow$
cylindrical hole $\rightarrow$ spherical hole $\rightarrow$ uniform.
This feature is also reproduced by Thomas-Fermi calculations
\cite{williams,lassaut,oyamatsu}.

The phases with these exotic nuclear structures,
if they are realized in NS crusts or SN cores,
bring about many astrophysical consequences.
As for those to NS phenomena,
it is interesting to note the relevance of nonspherical nuclei
in neutron star matter (NSM)
to pulsar glitches \cite{pp} and cooling of NSs \cite{lorenz}.
For supernova matter (SNM),
``pasta'' phases are expected to affect
the neutrino transport \cite{rpw,gentaro2}
and hydrodynamics in SN cores \cite{lassaut}.

Though the properties of ``pasta'' phases in equilibrium state
have been investigated actively,
the formation and the melting processes of them have not been discussed
except for some limited cases
based on perturbative approaches \cite{review,iida}.
It is important to adopt a microscopic and dynamical approach
which allows arbitrary nuclear structures
to understand these processes of nonspherical nuclei.
At finite temperatures, it is considered that
not only nuclear surface becomes obscure but also nuclei of various shapes
may coexist.
Therefore, it is necessary to incorporate density fluctuations
without any assumptions on nuclear shape
to investigate the properties of ``pasta'' phases at finite temperatures.
Although some previous works \cite{williams,lassaut}
do not assume nuclear structure,
they can not incorporate fluctuations of nucleon distributions
satisfactorily
because they are based on the Thomas-Fermi calculation
which is one-body approximation.
In addition, only one structure is contained in the simulation box
in these works,
there are thus possibilities that nuclear shape is strongly affected
by boundary effect and some structures are
prohibited implicitly.

We have started studying dense matter at subnuclear densities
by quantum molecular dynamics (QMD) \cite{aichelin}
which is one of the molecular dynamics (MD) approaches
for nucleon many body systems (see, e.g., Ref. \cite{feldmeier} for review).
MD for nucleons including QMD,
which is a microscopic and dynamical method
without any assumptions on nuclear structure,
is suitable for incorporating fluctuations of particle distributions.
The final aim of our study is to understand
the above mentioned formation and melting processes of nonspherical nuclei
and to investigate the properties of matter consists of nonspherical nuclei
at finite temperatures.
For the first step, the question posed in this paper is
as follows: {\it Can the phases with nonspherical nuclei
be formed dynamically?}

There is a pioneering work by Maruyama et al. \cite{maruyama}
which attempts to investigate the structure of matter
at subnuclear densities by QMD.
Unfortunately, they did not treat the Coulomb interaction consistently
and could not anneal the system successfully.
As a result, they could not reproduce the phases
with nonspherical nuclei of simple structures.
In the present work, we improve the above shortcomings and
obtain phase diagrams for cold dense matter of the proton fraction $x=0.3$
and 0.5 at subnuclear densities.
We also discuss nuclear shape changes with
morphological measures.

While there are a lot of versions of MD for fermions,
we choose QMD among them based on trade-off
between calculational amounts and accuracies.
The typical length scale $l$ of inter-structure
is $l \sim 10$ fm and the density region of interest is
just below the normal nuclear density $\rho_{0} = 0.165 {\rm\ fm}^{-3}$.
The required nucleon number $N$ in order to reproduce $n$ structures
in the simulation box is about $N \sim \rho_{0} (n l)^{3}$ (for slabs),
it is thus desirable that we prepare nucleons of order 10000
if we try to reduce boundary effects.
While it is very hard task to treat such a large system
by, for example, FMD and AMD (see, e.g., Ref. \cite{feldmeier} and
references therein)
whose calculational amounts scale as $\sim N^{4}$,
it is feasible to do it by QMD whose calculational amounts scale
as $\sim N^{2}$.
It is also noted that we mainly focus on macroscopic structures;
the exchange effect would not be so important for them.
Therefore, QMD, which is less elaborate
in treating the exchange effect,
has the advantage of other models.

We have performed QMD simulations of infinite (n,p,e) system
with fixed proton fraction $x = $ 0.3 and 0.5
for various nucleon densities $\rho$
( the density region is 0.05 - 1.0 $\rho_{0}$).
We set a cubic box which is imposed periodic boundary conditions
in which 2048 nucleons (1372 nucleons in some cases) are contained.
The relativistic degenerate electrons which ensure the charge neutrality
are treated as a uniform background
and the Coulomb interaction is calculated by the Ewald method
(see, e.g., Ref. \cite{allen})
which enables us to sum contributions of long-range interactions
in a system with periodic boundary conditions.
For nuclear interaction, we use an effective Hamiltonian
developed by Maruyama et al. (medium EOS model) \cite{maruyama}
which reproduce the bulk properties of nuclear matter
and the properties of stable nuclei, especially heavier ones,
i.e. binding energy and root-mean-square radius.

We first prepare an uniform hot nucleon gas
at $k_{B}T \sim 20$ MeV as an initial condition
which is equilibrated for $\sim 500 - 2000$ fm/c in advance.
In order to realize the ground state of matter,
we then cool it down slowly for $O(10^{3}-10^{4})$ fm/c
keeping the nucleon density constant
by frictional relaxation method etc.
until the temperature gets $\sim 0.1$ MeV or less.
Note that any artificial fluctuations are not given during the simulation.

The QMD equations of motion with friction terms
are solved using the fourth-order Gear predictor-corrector method
in conjunction with multiple time step algorithm \cite{allen}.
Integration time steps $\Delta t$ are set to be adaptive
in the range of $\Delta t < 0.1-0.2$ fm/c
depending on the degree of convergence.
At each step, the correcting operation is iterated until
the error of position $\Delta r$ and the relative error of momentum
$\Delta p/p$ get smaller than $10^{-6}$,
where $\Delta r$ and $\Delta p/p$ are estimated
as the maximum values of correction among all particles.
Computer systems which we use are equipped with MD-GRAPE II.

\begin{figure*}
\resizebox{15cm}{!}{\includegraphics{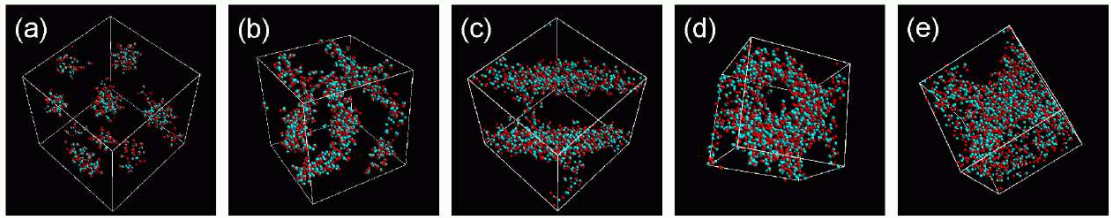}}
\caption{\label{pasta_sym}
  The nucleon distributions of typical phases with simple structures
  of cold matter at $x=0.5$;
  (a) sphere phase, $0.1 \rho_{0}$ ($D=43.65\ {\rm fm}$, $N=1372$);
  (b) cylinder phase, $0.18 \rho_{0}$ ($D=41.01\ {\rm fm}$, $N=2048$);
  (c) slab phase, $0.4 \rho_{0}$ ($D=31.42\ {\rm fm}$, $N=2048$);
  (d) cylindrical hole phase, $0.5 \rho_{0}$ ($D=29.17\ {\rm fm}$, $N=2048$) and
  (e) spherical hole phase, $0.6 \rho_{0}$ ($D=27.45\ {\rm fm}$, $N=2048$),
  where $D$ is the box size.
  The red particles represent protons and the green ones represent neutrons.}
\end{figure*}

\begin{figure*}
\resizebox{15cm}{!}{\includegraphics{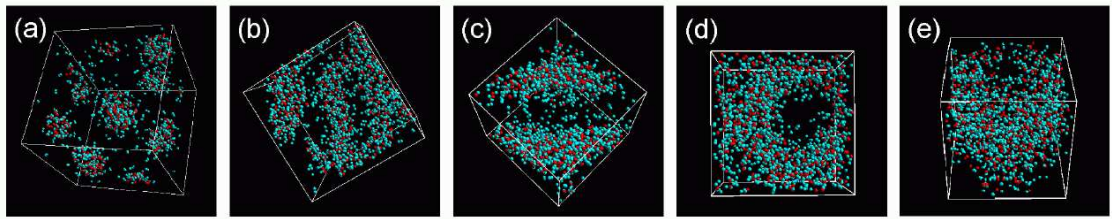}}%
\caption{\label{pasta_sym}
  Same as Fig.\ 1 at $x=0.3$;
  (a) sphere phase, $0.1 \rho_{0}$ ($D=49.88\ {\rm fm}$, $N=2048$);
  (b) cylinder phase, $0.18 \rho_{0}$ ($D=43.58\ {\rm fm}$, $N=2048$);
  (c) slab phase, $0.35 \rho_{0}$ ($D=32.85\ {\rm fm}$, $N=2048$);
  (d) cylindrical hole phase, $0.5 \rho_{0}$ ($D=29.17\ {\rm fm}$, $N=2048$) and
  (e) spherical hole phase, $0.55 \rho_{0}$ ($D=28.26\ {\rm fm}$, $N=2048$).
  The red particles represent protons and the green ones represent neutrons.}
\end{figure*}

Shown in Figs.\ 1 and 2 are the resultant nucleon distributions
of cold matter at $x=$ 0.5 and 0.3, respectively.
We can see from these figures that
the phases with rod-like and slab-like nuclei,
cylindrical and spherical bubbles,
in addition to the one with spherical nuclei are reproduced
in the both cases.
We here would like to mention some reasons of discrepancies
between the present result and the result obtained by Maruyama et al.
which says ``the nuclear shape may not have these simple symmetries''
\cite{maruyama}.
The most crucial reason seems to be the difference in
treatment of the Coulomb interaction.
In the present simulation,
we calculate the long range Coulomb interaction in a consistent way
using the Ewald method.
For the system of interest where the Thomas-Fermi screening length is
comparable to or larger than the size of nuclei,
this treatment is more adequate than that with introducing
an artificial cutoff distance as in Ref. \cite{maruyama}.
The second reason would be the difference
in the relaxation time scales $\tau$.
In our simulation, we can reproduce the bubble-phases
(see d and e of Figs. 1 and 2)
with $\tau \sim 10^{3}$ fm/c
and the nucleus-phases
(see b and c of Figs. 1 and 2)
with $\tau \sim O(10^{4})$ fm/c.
However, the matter in the density region
corresponding to a nucleus-phase
is quenched in an amorphous state when $\tau \alt 10^{3}$ fm/c.
In the present work, we take $\tau$ much larger than
typical time scale $\tau_{th} \sim O(100)$ fm/c
for nucleons to thermally diffuse
the distance of $l \sim 10$ fm at $\rho \simeq \rho_{0}$ and
$k_{B}T \simeq 1$ MeV.
This temperature is below the typical value of
the liquid-gas phase transition temperature
in the density region of interest,
it is thus considered that our results are thermally relaxed
in a satisfying level.

\begin{figure}
\rotatebox{270}{
\resizebox{5cm}{!}
{\includegraphics{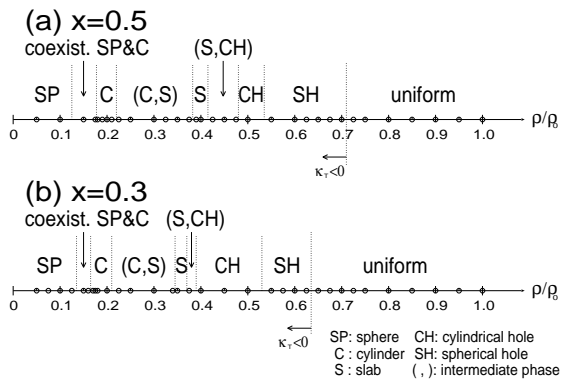}}}%
\caption{\label{pasta_sym}
  Phase diagrams of cold matter at $x=0.5$ (a) and $x=0.3$ (b).
  Matter is unstable against phase separation in the density region
  shown as $\kappa_{\rm T} < 0$,
  where $\kappa_{\rm T}$ is the isothermal compressibility.
  The symbols SP, C, S, CH and SH stand for nuclear shapes,
  i.e. sphere, cylinder, slab, cylindrical hole and spherical hole,
  respectively.
  The parentheses (A,B) show intermediate phases between A-phase and B-phase
  suggested in this work.
  They have complicated structures different from those of
  both A-phase and B-phase.
  Simulations have been carried out at densities denoted by small circles.}
\end{figure}

Phase diagrams of matter in the ground state are shown in Figs.\ 3 (a) and (b)
for $x=$ 0.5 and 0.3, respectively.
As can be seen from these figures,
the obtained phase diagrams basically reproduce
the sequence of the energetically favored nuclear shapes
predicted by simple discussions \cite{hashimoto}
which only take account of the Coulomb and surface effects;
this prediction is that the nuclear shape changes like
\ sphere $\rightarrow$ cylinder $\rightarrow$ slab $\rightarrow$
cylindrical hole $\rightarrow$ spherical hole $\rightarrow$ uniform,
with increasing density.
Comparing Figs.\ 3 (a) and (b),
we can see that the phase diagram shifts towards the lower density side
with decreasing $x$, which is due to the tendency
that as the nuclear matter at larger neutron excess,
the saturation density is lowered.
It is remarkable that the density dependence of the nuclear shape,
except for cylindrical bubbles (just in the case of $x=0.3$) and
spherical nuclei and bubbles, is quite sensitive
and phases with intermediate nuclear shapes
which are not simple as shown in Figs. 1 and 2
are observed in two density regions:
one is between the cylinder phase and the slab phase,
the other is between the slab phase and the cylindrical hole phase.
We note that these phases are different from coexistence phases
with nuclei of simple shapes and we will referred to them as
``intermediate phases''.

To extract the morphological characteristics of the nuclear shape changes
and the intermediate phases,
we introduce the Minkowski functionals
(see, e.g., Ref. \cite{minkowski} and references therein)
as geometrical and topological measures of the nuclear surface.
Let us consider a homogeneous body $K \in {\cal R}$
in the $d$-dimensional Eucledian space,
where ${\cal R}$ is the class of such bodies.
Morphological measures are defined as functionals
$\varphi: {\cal R} \rightarrow {\bf R}$
which satisfy the following three
properties:
(1) {\it Motion invariance}, i.e. $\varphi (K) = \varphi (gK)$,
where $g$ denotes any translations and rotations.
(2) {\it Additivity}, i.e.
$\varphi (K_{1} \cup K_{2})
= \varphi(K_{1})+\varphi(K_{2})-\varphi(K_{1} \cap K_{2})$,
where $K_{1},K_{2} \in {\cal R}$.
(3) {\it Continuity}, i.e. $lim_{n\to\infty} \varphi(K_{n}) = \varphi(K)$
if $\ lim_{n\to\infty} K_{n}=K$, where $K$ is a convex body and
$\{ K_{n} \}$ is a sequence of convex bodies.
Hadwiger's theorem
states that
there are just $d+1$ independent functionals
which satisfy the above properties;
they are known as Minkowski functionals.
In three dimensional space, four Minkowski functionals are related to
the volume, the surface area, the integral mean curvature
and the Euler characteristic.

Here, we particularly focus on the integral mean curvature
and the Euler characteristic; the results of other quantities
will be discussed elsewhere.
Both are described by surface integrals of the following local quantities,
the mean curvature $H = (\kappa_{1}+\kappa_{2})/2$ and
the Gaussian curvature $G = \kappa_{1} \kappa_{2}$,
i.e. $\int_{\partial K} H dA$ and
$\chi \equiv \frac{1}{2 \pi} \int_{\partial K} G dA$,
where $\kappa_{1}$ and $\kappa_{2}$ are the principal curvatures and
$dA$ is the area element of the surface of the body $K$.
The Euler characteristic $\chi$ is a purely topological quantity and
\begin{eqnarray}
  \chi & = & \mbox{(number of isolated regions)}
  - \mbox{(number of tunnels)} \nonumber\\
  && + \mbox{(number of cavities)}.
  \label{euler}
\end{eqnarray}
Thus $\chi > 0$ for the sphere and the spherical hole phases
and the coexistence phase of spheres and cylinders,
and $\chi = 0$ for the other ideal ``pasta'' phases,
i.e. the cylinder, the slab and the cylindrical hole phases.
We introduce the area-averaged mean curvature,
$\langle H \rangle \equiv \frac{1}{A}\int H dA$,
and the Euler characteristic density, $\chi / V$, as normalized quantities,
where $V$ is the volume of the whole space.

We calculate these quantities by the following procedure.
We first construct proton and nucleon density distributions
$\rho_{\rm p} ({\bf r}) = \left| \Phi_{\rm p} ({\bf r}) \right|^{2}$ and
$\rho ({\bf r}) = \left| \Phi ({\bf r}) \right|^{2}$
from data of the centers of position of the nucleons,
where $\Phi_{\rm p} ({\bf r})$ and $\Phi ({\bf r})$
are the QMD trial wave functions of protons and nucleons
(see Ref. \cite{maruyama}).
We set a threshold proton density $\rho_{\rm p, th}$ and then calculate
$f(\rho_{\rm p, th}) \equiv V(\rho_{\rm p, th})/A(\rho_{\rm p, th})$,
where $V(\rho_{\rm p, th})$ and $A(\rho_{\rm p, th})$ are
the volume and the surface area of the regions
in which $\rho_{\rm p}({\bf r}) \geq \rho_{\rm p, th}$.
We find out the value $\rho_{\rm p, th} = \rho_{\rm p, th}^{*}$
where $\frac{d^{2}}{d \rho_{\rm p, th}^{2}} f(\rho_{\rm p, th}^{*}) = 0$
and define the regions in which
$\rho_{\rm p}({\bf r}) \geq \rho_{\rm p, th}^{*}$ as nuclear regions.
For spherical nuclei, for example, $\rho_{\rm p, th}^{*}$
corresponds to a point of inflection of a radial density distribution.
In the almost whole phase-separating region,
the values of $\rho_{\rm th}^{*}$ distribute
in the range of about $0.7-0.9 \rho_{0}$
in the both cases of $x = 0.5$ and 0.3,
where $\rho_{\rm th}^{*}$ is the threshold nucleon density
corresponds to $\rho_{\rm p, th}^{*}$.
We then calculate $A$, $\int H dA$ and $\chi$
for the determined nuclear surface.
We evaluate $A$ by the triangle decomposition method,
$\int H dA$ by the algorithm shown in Ref. \cite{minkowski}
in conjunction with a calibration by correction of surface area,
and $\chi$ by the algorithm of Ref. \cite{minkowski} and
by that of counting deficit angles \cite{genus}
which are confirmed that both of them give the same results.

\begin{figure}
\rotatebox{0}{
\resizebox{8.2cm}{!}
{\includegraphics{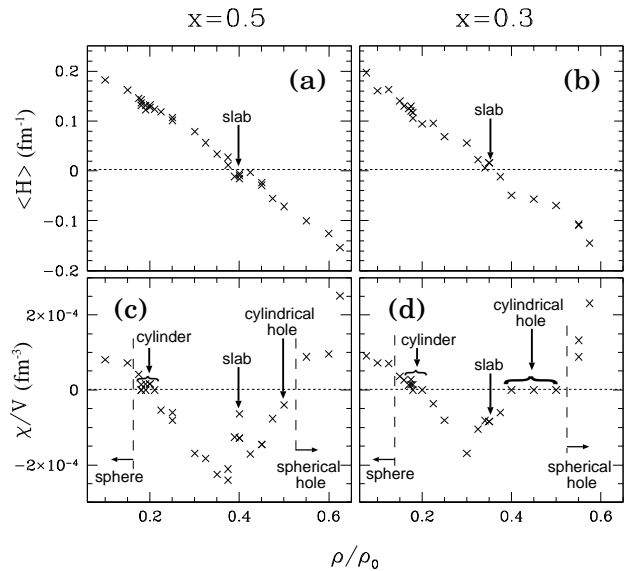}}}%
\caption{\label{pasta_sym}
  Density dependence of the Minkowski functionals
  of cold matter at $x=0.5$ (a and c) and
  $x=0.3$ (b and d).}
\end{figure}

We have plotted the obtained $\rho$ dependence of $\langle H \rangle$
and of $\chi / V$ for the surface of $\rho_{th}=\rho_{th}^{*}$
in Fig. 4.
In addition to the values of $\langle H \rangle$
for the surface of $\rho_{th}=\rho_{th}^{*}$,
we have also investigated those for the surface of
$\rho_{th}=\rho_{th}^{*} \pm 0.05\rho_{0}$
to examine the extent of the uncertainties of this quantity
which stem from the arbitrariness in the definition of the nuclear surface,
but we could not observe remarkable differences
from the values for $\rho_{th}=\rho_{th}^{*}$
(they were smaller than 0.015/fm).
We could not see these kinds of uncertainties in $\chi / V$
except for the densities near below the density
at which matter turns into uniform.

The behavior of $\langle H \rangle$ shows that
it decreases almost monotonously from positive to negative
with increasing $\rho$ until the matter turns into uniform.
The densities correspond to $\langle H \rangle \simeq 0$
are about 0.4 and 0.35$\rho_{0}$ for $x=0.5$ and 0.3, respectively;
these values are consistent with the density regions
of the phase with slab-like nuclei (see Fig. 3).
As mentioned previously, $\chi / V$ is actually positive
in the density regions corresponding to the phases with spherical nuclei,
coexistence of spherical and cylindrical nuclei, and spherical holes
because of the existence of isolated regions.
As for those corresponding to the phases with cylindrical nuclei,
planar nuclei and cylindrical holes, $\chi / V \simeq 0$.
The fact that the values of $\chi / V$ are not exactly zero
for nucleon distributions shown as slab phases in Figs. 1 and 2
reflects the imperfection of these ``slabs'',
which is due to the small nuclear parts which connect the neighboring slabs.
However, we can say that the behavior of $\chi / V$
depicted in Figs. 4(c) and 4(d) shows that
$\chi / V$ is negative in the density regions of the intermediate phases,
even if we take account of the imperfection of the obtained nuclear shapes
and the uncertainties of the definition of the nuclear surface.
This means that the intermediate phases consist of nuclear surfaces
which are saddle-like at each point on average
and they consist of each highly connected nuclear and gas regions
due to a lot of tunnels (see Eq. (\ref{euler})).

Let us now refer to discrepancies from the results of previous works
which do not assume nuclear structure \cite{williams,lassaut};
the intermediate phases can not be seen in these works.
We can give following two reasons for the discrepancies:
(1) These previous calculations are based on the Thomas-Fermi approximation
which can not sufficiently incorporate fluctuations of nucleon distributions.
This shortcoming may result in favoring nuclei of smoothed simple shapes
than in the real situation.
(2) There is a large possibility that some highly connected structures
which have two or more substructures in a period
are neglected in these works
because only one structure is contained in a simulation box.

If the phases with highly connected nuclear and bubble regions are realized
as the most energetically stable state,
we can say that it is not unnatural thing \cite{magierski}.
It is considered that, for example,
a phase with perforated slab-like nuclei, which has negative $\chi / V$,
could be more energetically stable than that with
extremely thin slab-like nuclei.
The thin planar nucleus costs surface-surface energy
which stems from the fact that nucleons in it feel surfaces of both sides.
We have to examine the existence of the intermediate phases
by more extensive simulations with larger nucleon numbers
and with longer relaxation time scales in the future.

Here we would like to discuss astrophysical consequences of our results.
Pethick and Potekhin have pointed out that
``pasta'' phases with rod-like and slab-like nuclei
are analogous to the liquid crystals
according to the similarity of the geometrical structures \cite{pp}.
It can also be said that the intermediate phases
observed in the present work are ``sponge''-like phases
because they have both highly connected nuclear and bubble regions
shown as $\chi / V < 0$.
The elastic properties of the sponge-like intermediate phases
are qualitatively different from
those of the liquid crystal-like ``pasta'' phases
because the former ones do not have any directions
in which restoring force does not act,
on the other hand the latter ones have.
Our results suggest that the intermediate phases
occupy a significant fraction of the density region
in which nonspherical nuclei can be seen (see Fig. 3).
If this is also true for more neutron-rich case as $x \sim 0.1$,
it leads to  increasing of the maximum elastic energy
that can be stored in the NS crust
than that in the case of the all nonspherical nuclei have simple structures.
Besides, the cylinder and the slab phases
which are liquid crystal-like lie between
the sponge-like intermediate phases or the crystalline solid-like phase,
and the releasing of the strain energy would, in consequence,
concentrate in the domain of these liquid crystal-like phases.
The above mentioned effects of the intermediate phases
should be taken into account
in considering the crust dynamics of starquakes etc.
if these phases exist in NSM.
In the context of pulsar glitch phenomena,
the effects of the sponge-like nuclei
on pinning rate of superfluid neutron vortices
also have yet to be investigated.

For neutrino cooling of NSs,
some version of the direct URCA process
which is suggested by Lorenz et al. \cite{lorenz}
that this might be allowed in the ``pasta'' phases
would be suppressed in the intermediate phases.
This is due to the fact that the proton spectrum is no longer continuous
in the sponge-like nuclei.
The last point which we would like to mention is
about the effects of the intermediate phases on neutrino trapping
in SN cores.
The nuclear parts connect over a wide region
which is much larger than that
characterized by the typical neutrino wave length $\sim 20$ fm.
Thus the neutrino scattering processes are no longer coherent
in contrast to the case of the spherical nuclei,
and this may, in consequence, reduce the diffusion time scale of neutrinos
as in the case of ``pasta'' phases with simple structures.
This reduction softens the SNM
and would thus act to enhance the amount of the released gravitational energy.

Our calculations demonstrate that the ``pasta'' phases
can be formed dynamically from hot uniform matter
in the proton-rich cases of $x=0.5$ and 0.3
without any assumptions on nuclear shape.
This suggests that the existence of these phases in NS crusts
because they cool down keeping the local thermal equilibrium
after proto-NSs are formed
and their cooling time scale is much larger than
the relaxation time scale of our simulations.
This conclusion has to be confirmed in more neutron-rich cases
of $x \sim 0.1$ in the future.
Our results also suggest that the existence of
the highly connected intermediate phases
which are characterized as $\chi / V < 0$.
This provides a vivid picture that
NS inner crusts which consist of dense matter at subnuclear densities
may be rich in properties
due to the possibilities of a variety of material phases.

\begin{acknowledgments}
G. W. are grateful to T. Maruyama, K. Iida,
K. Niita, A. Tohsaki, K. Oyamatsu,
S. Chikazumi, C. Hikage and K. Kotake
for helpful discussions and comments.
This work was supported in part
by the Junior Research Associate Program in RIKEN
through Research Grant No. J130026 and
by Grants-in-Aid for Scientific Research
provided by the Ministry of
Education, Culture, Sports, Science and Technology
through Research Grant No. 14102004 and No. 14-7939.
\end{acknowledgments}


\end{document}